\documentclass[aps,prl,twocolumn,showpacs]{revtex4}

\usepackage{graphicx}

\begin{document}

\title{
Pseudogap, Superconducting Energy Scale, and Fermi Arcs in
Underdoped Cuprate Superconductors }

\author{Hai-Hu Wen$^1$, Lei Shan$^1$, Xiao-Gang Wen$^2$, Yue Wang$^1$, Hong Gao$^1$, Zhi-Yong Liu$^1$, Fang Zhou$^1$, Jiwu Xiong$^1$, Wenxin Ti$^1$}

\affiliation{$^1$National Laboratory for Superconductivity,
Institute of Physics, Chinese Academy of Sciences, P.~O.~Box 603,
Beijing 100080, P.~R.~China}

\affiliation{$^2$Department of Physics, Massachusetts Institute of
Technology, Cambridge, Massachusetts 02139, USA}

\date{\today}

\begin{abstract}
Through the measurements of magnetic field dependence of specific
heat in $La_{2-x}Sr_xCuO_4$ in zero temperature limit, we
determined the nodal slope $v_\Delta$ of the quasiparticle gap. It
is found that $v_\Delta$ has a very similar doping dependence of
the pseudogap temperature $T^*$ or value $\Delta_p$. Meanwhile the
virtual maximum gap at ($\pi,0$) derived from $v_\Delta$ is found
to follow the simple relation $\Delta_q=0.46k_BT^*$ upon changing
the doping concentration. This strongly suggests a close
relationship between the pseudogap and superconductivity. It is
further found that the superconducting transition temperature is
determined by both the residual density of states of the pseudogap
phase and the nodal gap slope in the zero temperature limit,
namely, $T_c \approx \beta v_\Delta \gamma_n(0)$, where
$\gamma_n(0)$ is the extracted zero temperature value of the
normal state specific heat coefficient which is proportional to
the size of the residual Fermi arc $k_{arc}$. This manifests that
the superconductivity may be formed by forming a new gap on the
Fermi arcs near nodes below $T_c$. These observations mimic the
key predictions of the SU(2) slave boson theory based on the
general resonating-valence-bond (RVB) picture.
\end{abstract}
 \maketitle

\section{Introduction}

Since the discovery of the cuprate superconductors, 19 years have
elapsed without a consensus about its mechanism. Many exotic
features beyond the Bardeen-Cooper-Schrieffer theory have been
observed. One of them is the observation of a pseudogap in the
electron spectral function near the antinodal points ($\pi$,0) and
(0,$\pi$) at a temperature $T^*>>T_c$\cite{Pseudogap}. In a
conventional BCS superconductor, this gapping process occurs
simultaneously with the superconductivity at $T_c$. It has been
heavily debated about the relationship between the pseudogap and
the superconductivity in cuprates. One scenario assumes that the
pseudogap $\Delta_p$ marks only a competing or coexisting order
with the superconductivity and it has nothing to do with the
pairing origin. However another picture, namely the Anderson's
resonating-valence-bond (RVB)\cite{RVB} model (and its
offspring)\cite{Prepair} predict that the spin-singlet pairing in
the RVB state (which causes the formation of the pseudogap) may
lend its pairing strength to the mobile electrons and make them to
naturally pair and then to condense at $T_c$. According to this
picture there should be a close relationship between the pseudogap
and the superconductivity.

\begin{figure}
\includegraphics[width=8cm]{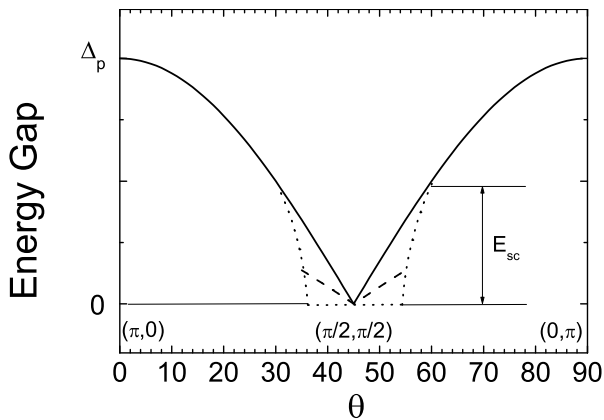}
\caption{Schematic plot for the pseudogap energy, superconducting
energy scale and nodal gap slope. The solid line represents a
standard d-wave gap $\Delta=\Delta_p cos2\theta$ with maximum gap
value $\Delta_p$ near ($\pi,0$) and $\theta$ the angle starting
from $k_x$. The dotted line shows the pseudogap near nodes if the
superconductivity would be suppressed completely ( based on the
Fermi arc picture ). The dashed line shows a possible
quasiparticle gap near nodes. The nodal gap slope is defined
as $v_\Delta=[d\Delta_s/d\theta]_{node}/\hbar k_F$.
The nodal gap slope $v_\Delta$ and the maximum gap $\Delta_p$
near ($\pi,0$) may not be related
if the superconductivity (which controls the gap
structure near nodes) has nothing to do with the pseudogap.}
\label{fig1}
\end{figure}

In order to check whether this basic idea is correct, we need to
collect the information for the pseudogap and the superconducting
energy scale, especially their doping dependence. The pseudogap
values $\Delta_p$ (or its corresponding temperature $k_BT^* \sim
\Delta_p$) and its doping dependence have been measured through
experiments\cite{Pseudogap}. To determine the superconducting
energy scale, we note that the normal state Fermi surface is
formed by four small arcs near the nodal points
\cite{PseudogapArc}.  As temperature is lowered below $T_c$, a new
gap opens on these arcs.  To illustrate this point more clearly,
in Fig.1 we present a schematic plot for different gaps or energy
scales.  The dotted line represents the gap structure of the
normal state, assuming the presence of Fermi arcs near the nodal
points.  The region of zero gap corresponds to the Fermi arc.  The
dash line and the solid line represent two possible gap structures
for superconducting state at $T=0$.  The solid line is the
standard d-wave gap with maximum gap value $\Delta_p$ at ($\pi,0$)
and ($0,\pi$).  From this picture, we see that the nodal gap
slope, which is defined as
$v_\Delta=[d\Delta_s/d\theta]_{node}/\hbar k_F$, can be used to
determine the superconducting energy scale.  The relationship
between the nodal gap slope $v_\Delta$ and the maximum pseudogap
$\Delta_p$ remains to be a big puzzle. In particular, the two
quantities may be independent of each other if the
superconductivity is not induced by the formation of the
pseudogap. Therefore to measure the nodal gap slope near nodal
point in the zero temperature limit becomes highly desired. When
combined with the known results on the pseudogap $\Delta_p$, this
will allow us to detect the relation between the pseudogap and the
superconductivity.

Some previous results using, for example, angle-resolved
photo-emission (ARPES)\cite{Mesot} or superfluid density seem to
be inconclusive due to either energy resolution (ARPES above
10meV) or unexpected difficulty in analyzing the data (e.g., a
so-called Fermi liquid correction factor $\alpha_{FL}$ is
inevitably involved in analyzing the low temperature data of
superfluid density). In this paper, we report the evidence of a
proportionality between the nodal gap slope $v_\Delta$ and the
pseudogap temperature $T^*$. Remarkably a simple relation, namely
$\Delta_q=0.46 T^*$, between the virtual maximum quasiparticle gap
($\Delta_q$) derived from $v_\Delta$ and the pseudogap temperature
$T^*$ is found. We also find that $T_c$ is determined by both the
nodal gap slope $v_\Delta$ and the size of the Fermi arcs
($k_{arc}$) in the underdoped normal state. Both observations are
anticipated by the SU(2) slave boson theory\cite{SU2} based on the
general RVB picture.

\section{Experiment}

We determine the properties of the nodal quasiparticles by
measuring the low temperature electronic specific heat. The
$La_{2-x}Sr_xCuO_4$ single crystals measured in this work were
prepared by travelling solvent floating-zone technique. Samples
with seven different doping concentrations p=0.063($T_c$=9K,
nominal x=0.063, post-annealed in $Ar$ gas at $800^\circ C$ for 48
hrs ), 0.069($T_c$=12K, as-grown sample with x=0.063), 0.075
($T_c$=15.6K, nominal x=0.07 and post-annealed in $O_2$ gas at
$750 ^\circ C$ for 12 hrs), 0.09 ($T_c$=24.4K, as grown, x=0.09),
0.11 ($T_c$=29.3K, as grown, x=0.11), 0.15 ($T_c$=36.1K, nominal
x=0.15), 0.22 ($T_c$=27.4K, nominal x=0.22) have been
investigated. The quality of our samples has been characterized by
x-ray diffraction, and $R(T)$ data showing a narrow transition
$\Delta T_c \leq $ 2 K. The samples have also been checked by AC
and DC magnetization showing also quite narrow transitions. The
full squares in Fig.6 represent the transition temperatures of our
samples. The heat capacity presented here was measured with the
relaxation method based on an Oxford cryogenic system
Maglab-EXA-12. In all measurements the magnetic field was applied
parallel to c-axis. As also observed by other groups for $La-214$
system, the anomalous upturn of $C/T$ due to the Schottky anomaly
of free spins is very weak. This avoids the complexity in the data
analysis. Details about the sample characterization, the specific
heat measurement, the residual linear term and extensive analysis
are reported in a recent paper\cite{WenPRB2004}.

\begin{figure}
\includegraphics[width=8cm]{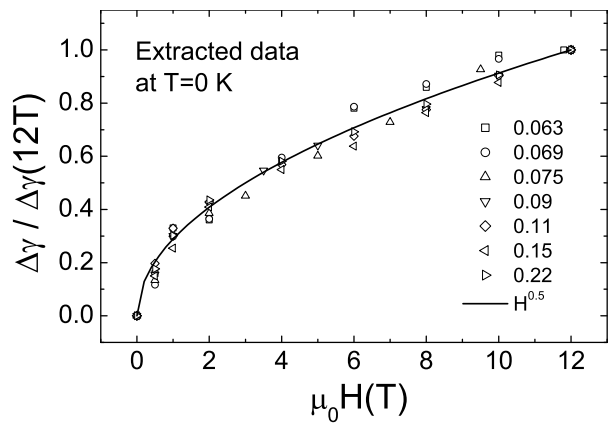}
\caption{Field dependence of $\Delta \gamma =[C(H)-C(0)]/T$
normalized by the data at about 12 T in zero temperature limit. It
is clear that Volovik's $\sqrt{H}$ relation describes the data
rather well for all samples. This indicates a robust $d-wave$
superconductivity in all doping regimes.} \label{fig2}
\end{figure}

\begin{figure}
\includegraphics[width=8cm]{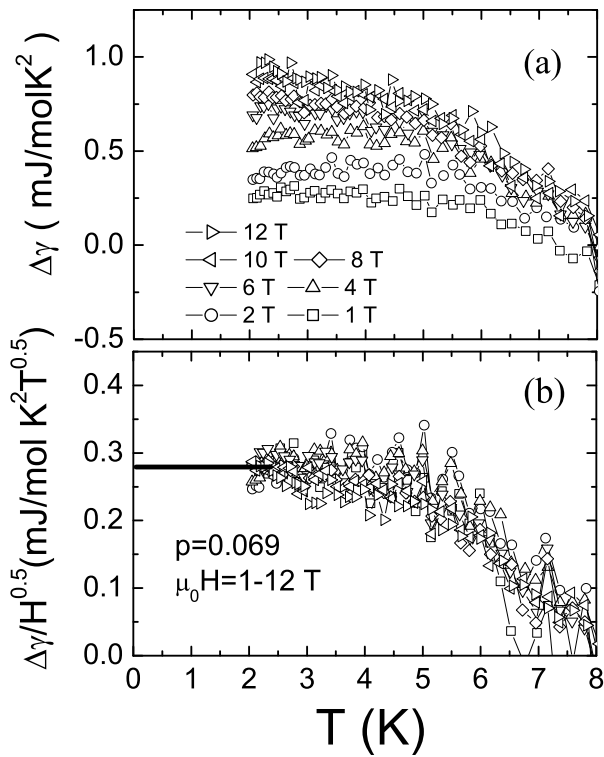}
\caption{(a) The typical original data of $\Delta\gamma$ vs. $T$
for the underdoped sample $p=0.069$ at different magnetic fields.
(b) The same set of data plotted as $\Delta\gamma/\sqrt{H}$ vs.
$T$. One can clearly see that in zero temperature limit
$\Delta\gamma / \sqrt{H}$ is a constant for all fields implying
the validity of the Volovik's relation $\Delta\gamma=A\sqrt{H}$.
From here one can also determine the value $A$ which is about
$0.28 mJ/mol K^2T^{0.5}$ as marked by the thick bar.} \label{fig3}
\end{figure}

It has been widely perceived that the pairing symmetry in the hole
doped cuprate superconductors is of d-wave with line nodes in the
gap function. In the mixed state, due to the presence of vortices,
Volovik \cite{Volovik} pointed out that supercurrents around a
vortex core lead to a Doppler shift to the quasi-particle
excitation spectrum. This will dominate the low energy excitation
and the specific heat (per mol) behaves
as\cite{Volovik,Kopnin1996} $C_{vol}=A\sqrt{H}$ with $A \propto
1/v_\Delta$. This square-root relation has been verified by many
measurements which were taken as evidence for d-wave symmetry, for
example by specific
heat\cite{Moler,Revaz,Wright,Phillips,Nohara,Chen,Hussey}, thermal
conductivity\cite{Taillefer}, tunnelling (to measure the Doppler
shift of the Andreev bound states)\cite{Greene}, etc. In this way
one can determine the nodal gap slope ($v_{\Delta}$). Since the
phonon part of the specific heat is independent on the magnetic
field, this allows to remove the phonon contribution by
subtracting the C/T at a certain field with that at zero field,
one has $\Delta\gamma=\Delta C/T=[C(H)-C(0)]/T=C_{vol}/T-\alpha T$
with $\alpha$ the coefficient for the quasiparticle excitations of
a d-wave superconductor at zero field ($C_e=\alpha T^2$). In the
zero temperature limit $\Delta \gamma=C_{vol}/T=A\sqrt{H}$ is
anticipated.

\section{Results and discussion}

In order to get $\Delta \gamma$ in the zero temperature limit, we
extrapolate the low temperature data of $C/T$ vs. $T^2$ (between
2K to 4K) to zero K. The data taken in this way and normalized at
12 T are presented in Fig.2. It should be mentioned that the
similar data have been published in our previous
paper\cite{WenPRB2004}. Here for clarity we present the data again
with more detailed analysis. It is clear that the Volovik's
$\sqrt{H}$ relation describes the data rather well for all doping
concentrations. This is to our surprise since it has been
questioned whether the Volovik relation is still obeyed in the
underdoped regime\cite{Nohara} especially when competing orders
are expected to appear\cite{SDW,DDW,AFSO5} and impurity scattering
is present. We attribute the success of using the Volovik relation
here to three reasons: (1) We use $\Delta \gamma=[C_{H
||c}-C_{H=0}]/T$ instead of using $\Delta \gamma=[C_{H
||c}-C_{H\perp c}]/T$. The latter may inevitably involve the
unknown DOS contributions from other kinds of vortices (for
example, Josephson vortices) when $H\perp C$. (2) The contribution
from a second competing order to $\Delta \gamma$ may be small
compared to the Volovik's term in the zero temperature limit. This
is reasonable when considering a contribution to the heat capacity
by the competing order as $\sim T^\nu$ with $\nu>1$. For example,
the specific heat due to the spin correlation in 2D
anti-ferromagnetic phase is $\sim T^2$. In zero temperature limit
this term goes away. (3) The DOS induced by the Doppler shift
effect in our experiment is much stronger than that induced by the
impurity scattering. We will further address this point in the
forthcoming discussion. To have a self-consistent check of the
$\sqrt{H}$ relation found in the zero temperature limit, we plot
the raw data of $\Delta\gamma/\sqrt{H}$ vs. $T$ at finite
temperatures. A typical example for the very underdoped one
($p=0.069$) is shown in Fig.3(a) and (b). One can see that in the
low temperature region the data $\Delta\gamma/\sqrt{H}$ scale for
all fields ranging from 1 T to 12 T, showing the nice consistency
with the relationship $\Delta\gamma \propto \sqrt{H}$ for this
sample in the zero temperature limit. From here one can also
determine the prefactor $A$ in $\Delta\gamma=A\sqrt{H}$ (here for
example, $A = 0.28 mJ/mol K^2T^{0.5}$ for p=0.069) and then
compare backwards to the value determined from the data shown in
the main panel leading to of course the same value. The same
feature appears for all other doping concentrations. For clarity
they will not be shown here.

\begin{figure}
\includegraphics[width=8cm]{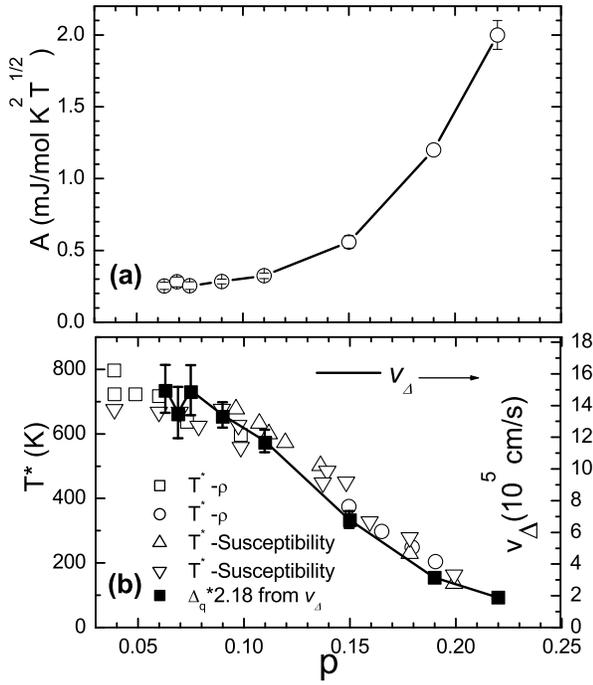}
\caption {(a) Doping dependence of the pre-factor $A$ determined
in present work (open circles). Here the point at p=0.19 was
adopted from the work by Nohara et al. on a single
crystal\cite{Nohara}. (b) Doping dependence of the pseudogap
temperature $T^*$ (open symbols) summarized in Ref.1 (see Fig.26
there) and our data $v_\Delta$ (solid line). $T^*$-susceptibility
refers to the pseudogap temperature determined from the maxima in
the static susceptibility, and $T^*-\rho$ to the temperature at
which there is a slope change in the DC resistivity (all in
$La_{2-x}Sr_xCuO_4$). Above $T^*-\rho$ the resistivity has a
linear temperature dependence. The full squares represent the
calculated virtual maximum quasi-particle gap $\Delta_q$ derived
from $v_\Delta$ without any adjusting parameters. Surprisingly
both set of data are correlated through a simple relation
($\Delta_q\approx 0.46 k_BT^*$) although they are determined in
totally different experiments. This result implies a close
relationship between the pseudogap $\Delta_p$ and the nodal gap
slope $v_\Delta$. } \label{fig4}
\end{figure}

\begin{figure}
\includegraphics[width=8cm]{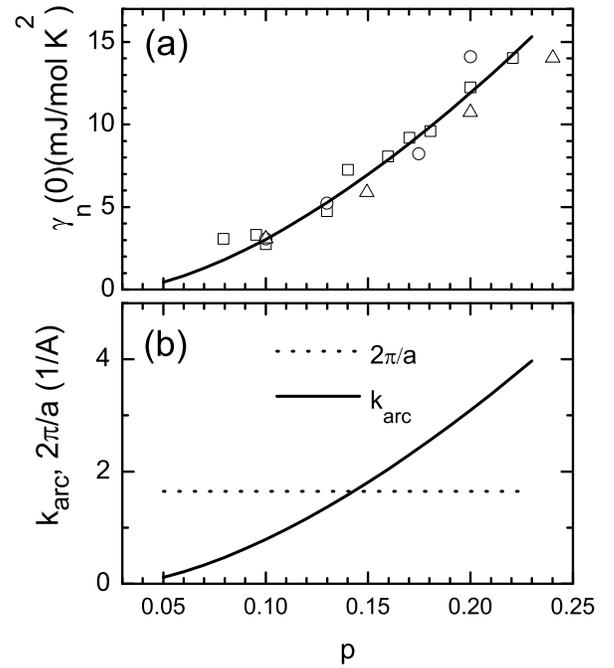}
\caption{(a)The doping dependence of $\gamma_n(0)$ derived from
specific heat (open squares\cite{Momono}) based on entropy
conservation, ARPES (open circles\cite{Ino}), and Knight shift
(up-triangles\cite{Ohsugi}) in $La_{2-x}Sr_xCuO_4$. The solid line
is a fit to the data with $\gamma_n=\zeta(p-p_c)^\eta$ yielding
$\zeta=182.6,p_c=0.03, \eta=1.54$. (b) A comparison between
$k_{arc}$ calculated from $\gamma_n(0)$ and $2\pi/a$. One can see
that the $k_{arc}$ becomes smaller than $2\pi/a$ in the underdoped
region showing the relevance of the Fermi arcs in the pseudogap
phase.} \label{fig5}
\end{figure}

It is clear that the Volovik's $\sqrt{H}$ relation describes the
data rather well for all doping concentrations. This successful
scaling of $\Delta \gamma$ vs. $\sqrt{H}$ makes it possible to
derive the pre-factor $A$, and one can further determine the nodal
gap slope $v_\Delta$. Fig.4(a) shows the doping dependence of the
pre-factor $A$. The error bar is obtained by fitting the extracted
zero temperature data to $\Delta \gamma = A \sqrt{H}$. For a
typical d-wave superconductor, by calculating the excitation
spectrum near the nodes, it was shown that \cite{Hussey}
\begin{equation}
\label{deltap}
 A=\alpha_p\frac{4k_B^2}{3\hbar l_c}\sqrt{\frac{\pi}{\Phi_0}}\frac{nV_{mol}}{v_\Delta}
\end{equation}
here $l_c$ = 13.28 $\AA$ is the c-axis lattice constant, $V_{mol}$
=58 $cm^3$ (the volume per mol), $\alpha_p$ a dimensionless
constant taking 0.5 (0.465) for a square (triangle) vortex
lattice, n=2 (the number of Cu-O plane in one unit cell), $\Phi_0$
the flux quanta. The $v_\Delta$ has then been calculated without
any adjusting parameter (taking $\alpha_p$=0.465) and shown in
Fig.4(b). It is remarkable that $v_\Delta$ has a very similar
doping dependence as the pseudogap temperature $T^*$, indicating
that $v_\Delta\propto T^*\propto \Delta_p$. If converting the data
$v_\Delta$ into the virtual maximum quasiparticle gap
($\Delta_q$)\cite{Hussey} via $v_\Delta=2\Delta_q/\hbar k_F$, here
$k_F\approx \pi/\sqrt{2}a$ is the Fermi vector of the nodal point
with $a=3.8\AA$ (the in-plane lattice constant), surprisingly the
resultant $\Delta_q$ value [shown by the filled squares in
Fig.4(b)] is related to $T^*$ in a simple way ($\Delta_q \sim 0.46
k_B T^*$). It is important to emphasize that this result is
obtained without any adjusting parameters. Counting the
uncertainties in determining $T^*$ and the value of $\alpha_p$,
this relation is remarkable since $\Delta_q$ and $T^*$ are
determined in totally different experiments. Because $v_\Delta$
(or $\Delta_q$) reflect mainly the information near nodes which is
predominantly contributed by the superconductivity, above
discovery, i.e., $v_\Delta\propto T^*\propto \Delta_p$ (or
$\Delta_q \sim 0.46 k_BT^*$) strongly suggests a close
relationship between the superconductivity and the pseudogap. A
similar conclusion was drawn in underdoped $YBa_2Cu_3O_y$ by
analyzing the low temperature thermal
conductivity\cite{Taillefer2}. If the pseudogap is supposed to be
caused by the formation of the RVB state\cite{RVB}, our results
here point to a fact that the RVB singlet pairing may be one of
the unavoidable ingredients for superconductivity. It remains to
know whether this conclusion holds also for the electron doped
samples since so far it is not clear yet whether the pseudogap
exists in these N-type samples.

In above discussion, we see the consistency between our low
temperature specific heat data and the Volovik's square root
relation $\Delta \gamma = A\sqrt{H}$. This seems surprising since
the temperature range we considered here is about several Kelvin.
At such an energy scale, the impurity scattering will strongly
alter the DOS in the low energy region by generating some new
quasiparticles. However, by applying a magnetic field, the Doppler
shift of the quasiparticle excitation spectrum will contribute a
new part to DOS. This energy shift is actually not small comparing
to the temperature. We can give a simple estimation on the energy
shift $\Delta E$. It is known that $\Delta E =
\alpha_{FL}\sqrt{\frac{\sqrt{3}}{2\pi}}{\frac{v_F\hbar}{l_B}}$,
here $l_B$ is the magnetic length which is defined as
$l_B=(c\hbar/eB)^{1/2}$, and $0<\alpha_{FL}<1$ is a Fermi liquid
correction term.  Taking $v_F=2.73\times 10^7cm/s$\cite{ZhouXJ},
we have $\Delta E=3.67\alpha_{FL}\sqrt{B/1T}$ $meV$.  For example,
taking the maximum field (12 T) in our experiment, we get $\Delta
E=12.2 \alpha_{FL}meV$ which is actually a relatively big energy
scale compared to the temperature $T$ since $\alpha_{FL}\sim 1$.
This may explain why the Volovik's simple square-root relation
$\Delta \gamma=A\sqrt{H}$ can be easily observed in our single
crystals with inevitable certain amount of impurities.

In the following we will investigate what determines $T_c$.
Bearing the doping dependence of $v_\Delta$ in mind, it is easy to
understand that $v_\Delta \hbar k_F$ should not be a good estimate
of the superconducting energy scale for the underdoped samples
since the $T_c$ and $v_\Delta$ have opposite doping dependence.
The basic reason is that the normal-state Fermi surfaces are small
arcs of length $k_{arc}$ near the nodal points. The
superconducting transition occurs by forming extra gaps on the
Fermi arcs. So the effective superconducting energy scale should
be estimated as $E_s \sim \frac{1}{2}v_\Delta \hbar k_{arc}$. From
the normal state electronic specific heat $C_{ele}=\gamma_nT$, we
have $\gamma_n=4nk_B^2k_{arc}V_{mol}/\hbar v_Fl_c$. Assuming $E_s
\sim k_B T_c$ we find
\begin{equation}
\label{Tc}
 T_c=\alpha_s \frac{\hbar^2 v_F l_c \gamma_n v_\Delta}{8n k_B^3
 V_{mol}}=\beta\gamma_nv_\Delta
\end{equation}
where $\alpha_s$ is a dimensionless constant in the order of
unity, $v_F$ is the nodal Fermi velocity normal the Fermi surface.
The value of $\gamma_n(0)$ can be estimated from specific
heat\cite{Momono,Why}, or indirectly by ARPES\cite{Ino} or
NMR\cite{Ohsugi}. Here we take the values for $\gamma_n(0)$
summarized by Matsuzaki et al.\cite{Momono} and fit it (in unit of
$mJ/mol K^2$) with a formula $\gamma_n=\zeta(p-p_c)^\eta$ yielding
$\zeta=182.6,p_c=0.03, \eta=1.54$. In Fig.5 we present the doping
dependence of the zero-temperature specific heat coefficient
$\gamma_n(0)$ and $k_{arc}$. One can see that $k_{arc}$ becomes
smaller than $2\pi/a$ in underdoped region showing the
self-consistency of the picture of Fermi arcs. In Fig.6 we present
the doping dependence of the truly measured $T_c$ (filled squares)
and the calculated value (open squares) by eq.(2) with
$\beta=0.7445 K^3mols/J m$. In underdoped region, the truly
measured and calculated $T_c$ values coincide rather well implying
the validity of eq.(2). In the overdoped region, $\gamma_n$ will
gradually become doping independent, therefore one expects
$T_c\propto v_\Delta$. So the energy scale of the
superconductivity is not given by $v_\Delta \hbar k_F\sim
\Delta_p$, but by $\frac{1}{2}v_\Delta \hbar k_{arc}$ or more
precisely by eq.(2) in the underdoped region.

\begin{figure}
\includegraphics[width=8cm]{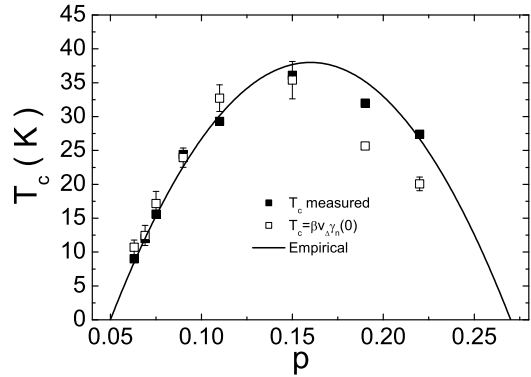}
\caption{ Doping dependence of the truly measured superconducting
transition temperature $T_c$ (full squares) and that calculated by
$T_c=\beta v_\Delta \gamma_n(0)$ (open squares) with $\beta=0.7445
K^3 mol s/J m$. The solid line represents the empirical relation
$T_c/T_c^{max}=1-82.6(p-0.16)^2$ with $T_c^{max}=38 K$.}
\label{fig6}
\end{figure}

To have a framework about the experimental results, in the
following, we will review one particular explanation based on the
slave-boson approach. Within the SU(2) slave-boson
theory\cite{SU2}, the pseudogap metallic state is viewed as a
doped algebraic spin liquid (ASL)\cite{ASL}. A doped ASL is
described by spinons (neutral spin-1/2 Dirac fermions) and holons
(spinless charge-e boson) coupled to a U(1) gauge field. Due to
the attraction between the spinons and the holons caused by the
U(1) gauge field, a spinon and a holon recombine into an electron
at low energies\cite{SU2,ASL}. Due to the spin-charge
recombination, the pseudogap metallic state is described by
electron-like quasiparticles at low energy. Since the binding
between the spinon and the holon is weak, the large pseudogap near
the anti-nodal points ($\pi$, 0) and (0, $\pi$) is not affected.
So the Fermi surface of the recombined electrons cannot form a
large closed loop. A simple theoretical calculation\cite{SU2}
suggests that the Fermi surface of the recombined electrons forms
four small arcs near the nodal points ($\pm \pi/2,\pm \pi/2$).
Thus the SU(2) slave boson theory\cite{SU2} contains two key
features: the pseudogap due to spin singlet pairing and the Fermi
arcs due to the spin-charge recombination\cite{ASL}. And the
superconductivity arises from the coherent motion of the
quasiparticles on the arcs, thus one expects that $T_c$ is
proportional to the gap on the Fermi arc: $k_BT_c\approx
\frac{1}{2}v_\Delta \hbar k_{arc}$, instead of the pseudogap
$\Delta_p$ near the anti-nodal points. Meanwhile, since the spin
pairing is responsible for both the pseudogap $\Delta_p$ near the
anti-nodal points and the nodal gap slope $v_\Delta$, it is
reasonable to see the proportionality between $v_\Delta$ and
$T^*(\propto \Delta_p)$ or $\Delta_q\approx 0.46 k_BT^*$. These
are exactly what we found in the experiment.

\section{Concluding Remarks}
In summary, the Volovik's relation of the d-wave pairing symmetry
has been well demonstrated by low temperature specific heat in
wide doping regime in $La_{2-x}Sr_xCuO_4$. Based on this analysis
the nodal gap slope $v_\Delta$ is derived and is found to follow
the same doping dependence of the pseudogap $\Delta_p$. This
strongly indicates the close relationship between the pseudogap
and the superconductivity. Meanwhile it is found that the
superconducting transition temperature $T_c$ is determined by
$v_\Delta \gamma_n(0)$ instead of $v_\Delta$. This discovery may
suggest the importance of Fermi arcs near the nodal region and the
superconductivity is induced by the formation of a new gap on
these arcs. Both observations are consistent with the SU(2) slave
boson theory based on the general RVB picture.

This work is supported by the National Science Foundation of
China, the Ministry of Science and Technology of China (973
project: 2006CB601002), the Knowledge Innovation Project of
Chinese Academy of Sciences. XGW is supported by NSF Grant
No.DMR--04--33632, NSF-MRSEC Grant No. DMR--02--13282, and NFSC
no. 10228408. We thank Yoichi Ando and his group (CRIEP, Komae,
Tokyo, Japan) for providing us some nice single crystals. We are
grateful for fruitful discussions with Dung-Hai Lee, Zhongxian
Zhao, Zhengyu Weng, Tao Xiang and Qianghua Wang.

Correspondence should be addressed to hhwen@aphy.iphy.ac.cn or
wen@dao.mit.edu.


\begin{thebibliography}{00}
\bibitem{PseudogapArc}Based on the entropy conservation, we would argue that the pseudogap is zero
near (not only at) the nodal point, i.e., a Fermi arc near nodal
point exists even at zero temperature in the pseudogap state. In
Ref.\cite{Momono}, one can see that the pure pseudogap phase (when
the superconductivity is completely suppressed) has actually a
finite DOS at zero K based on the entropy conservation. This
finite DOS at zero K of the pseudogap phase can be interpreted as
due to two possible reasons: one is inuced by the impurity
scattering of the nodal quasiparticles of a d-wave pseudogap,
another one is given by the zero-temperature Fermi arcs of the
pseudogap. However in our later discussion we will see that the
residual DOS at zero K in the pseudogap phase increases almost
linearly with $p$ up to the optimal doping point. This cannot be
understood by the picture of impurity scattering since it is known
that the optimally doped sample may have less disorders than the
underdoped one. Therefore it is tempting to conclude that there
are Fermi arcs near nodes in the pseudogap phase in the zero
temperature limit.
\bibitem{Why}In principle, $\gamma_n(0)$ can be determined from
ourown zero field specific heat data, however since the zero-field
term $\alpha T^2$ is too small compared to other terms (phonon,
residual linear term), this makes it almost impossible to get
relevant $\gamma_n(0)$ using our technique. Therefore here we use
the data derived based on the entropy conservation by Matsuzaki et
al.\cite{Momono}. The advantage of our relaxation techniqe is to
determine the field induced change of $\gamma$ in low temperature
region.
\end{thebibliography}
\end{document}